\documentclass[fleqn,twoside]{article}
\usepackage{espcrc2}
\usepackage{epsfig}

\usepackage{graphicx}
\usepackage[figuresright]{rotating}

\newcommand{\AmS}{{\protect\the\textfont2
  A\kern-.1667em\lower.5ex\hbox{M}\kern-.125emS}}

\title{Evidence for a BKT transition and a pseudogap phase in three-dimensional 
        Gross-Neveu model at $T \neq 0$}

\author{Costas G. Strouthos\address{Department of Physics,
        University of Wales Swansea, \\
        Singleton Park, Swansea SA2 8PP, United Kingdom}
        }

\begin{document}

\begin{abstract}
We present results from Monte Carlo simulations of the three-dimensional Gross-Neveu 
Model with a $U(1)$ chiral symmetry at nonzero temperature. We provide  evidence
that the model undergoes a Berezinskii-Kosterlitz-Thouless transition in accordance with 
the dimensional reduction scenario. We also identify 
a regime in the high temperature phase in which the fermions acquire nonzero dynamical mass, 
analogous to the pseudogap behaviour observed in cuprate superconductors. 

\vspace{1pc}
\end{abstract}

\maketitle

\section{Introduction}
The three-dimensional Gross-Neveu model (GNM$_3$) has been proved to be an interesting
and tractable model to study chiral phase transitions both numerically by means of lattice simulations
and analytically in the form of large-$N_f$ expansions.
In this report we present results of numerical simulations of the $U(N_f)_V$-invariant GNM$_3$ with a
$U(1)$ chiral symmetry at non-zero temperature. 
This model is described by the
following continuum Euclidean Lagrangian density:
\begin{equation}
{\cal L}= \bar{\Psi}_i(\partial\hskip -.5em /  + \sigma + i \gamma_5 \pi)\Psi_i
+ \frac{N_f}{2 g^{2}} (\sigma^{2}+ \pi^2).
\end{equation}
We treat $\Psi_i$, $\bar{\Psi}_i$  as four-component Dirac spinors and the index $i$ runs over $N_f$
fermion species.
The model is renormalizable in the $1/N_f$ expansion unlike in the loop
expansion \cite{rosen91}. For sufficiently strong coupling at $T\!=\!0$ 
it exhibits spontaneous chiral symmetry breaking and if we choose $\langle \pi \rangle =0$ 
the pion field $\pi$
is the associated Goldstone boson.

At leading order in $1/N_f$ this  model undergoes a second order chiral phase
transition \cite{rosen91}. 
%at $T_c = \frac{m_f}{2 \ln2}$ \cite{rosen91},
%where $m_f$ is the fermion dynamical mass at
%zero temperature.
%The leading order effective potential has the same form as the discrete symmetry case with the replacement
%$\sigma^2 \! \rightarrow \! \sigma^2 \! + \! \pi^2$.
This conclusion is expected to be valid only when $N_f$ is strictly infinite, i.e. when the fluctuations of the
bosonic fields are neglected, otherwise it runs foul of
the Coleman-Mermin-Wagner theorem, which states that in two-dimensional systems
the continuous
chiral symmetry must be manifest for all $T \! > \! 0$.
Next-to-leading order calculations \cite{babaev,appelquist} demonstrated that the 
the model undergoes a Berezinskii-Kosterlitz-Thouless (BKT) transition \cite{kosterlitz}
in accordance with the dimensional reduction scenario.
In \cite{strouthos98} it was shown that the thermal transition of the $Z_2$-symmetric GNM$_3$ model 
belongs to the $2\!-\!d$ Ising universality class.
%symmetry is restored for arbitrarily small
%$T \! < \! 0$ [7$-$10].
%As argued in \cite{appelquist,babaev}
%the model is expected in accordance with the dimensional reduction scenario
%to undergo a Berezinskii-Kosterlitz-Thouless (BKT) transition \cite{kosterlitz}
%at $T_{BKT}$ which is associated with the
%unbinding of vortices like in the two-dimensional $XY$ model.
%In terms of the reduced temperature $t \equiv T-T_{BKT}$ the scaling behaviour of the correlation
%length, susceptibility and the specific heat is given \cite{kosterlitz} by
%\begin{equation}
%\xi(t) \sim e^{at^{-\nu}}\!\!, \;\;\; \chi(t) \sim \xi^{2-\eta}\!\!, \;\;\; C_v \sim \xi^{\alpha/\nu}+\mathrm{const.},
%\end{equation}
%where for $t \rightarrow 0^+$, $\nu=1/2$, $\eta=1/4$ and $\alpha=-d\nu=-1$.
It is easier to visualize the BKT scenario if we use the ``modulus-phase'' parametrization
$\sigma + i \pi \equiv \rho e^{i \theta}$.
In two spatial dimensions logarithmically divergent infrared fluctuations do not allow the phase
$\theta$ to take a fixed direction and therefore prevent spontaneous symmetry
breaking via $\langle\theta\rangle\not=0$.
The critical temperature $T_{BKT}$ is expected to separate two \emph{different} chirally symmetric phases:
a low $T$ phase, which is characterized by power law phase correlations
$\langle e^{i\theta(x)} e^{-i\theta(0)} \rangle \sim x^{-\eta(T)}$ at distances $x \gg 1/T$
but no long range order (i.e.
a spinwave phase where chiral symmetry is ``almost but not quite broken''), and a high $T$
phase which is characterized by exponentially decaying phase correlations with no long range order.  In
other words for $0 \leq T \leq T_{BKT}$
there is a line of critical points characterised by a
continuously varying $0\leq\eta(T)\leq{1\over4}$.
However, since the amplitude $\rho$ is neutral under $U(1)$,
the dynamics of the low temperature phase do not preclude the
generation of a fermion mass $m_f\propto\rho$
whose value may be comparable with the naive
prediction of the large-$N_f$ approach. 
In \cite{babaev} it is
shown that in GNM$_3$ with $T\!>\!0$,
next-to-leading order corrections cause fermion mass generation
to occur for $T\!<\!T_*$, where $T_*\!>\!T_{BKT}$,
i.e. it predicts a ``pseudogap'' phase in which the system is non-critical.
In the next section we present numerical evidence in favour of these 
scenarios \cite{strouthos01}.

\section{Results}

The model was formulated on the lattice using the staggered fermion 
formulation and the simulations were performed using the standard 
hybrid Monte Carlo algorithm. Details concerning the lattice action 
and the algorithm can be found in \cite{hands93}.
\begin{figure}[t]
\centering
\includegraphics[scale=0.39]{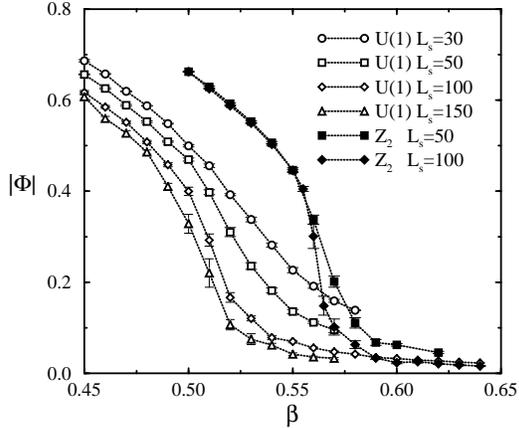}
\vspace{-1.3cm}
\caption{\small Order parameter $\vert\Phi\vert$ vs. $\beta$ for $L_t=4$
for both $U(1)$ and $Z_2$ symmetric models.}
\label{fig:order_para}
\end{figure}
%\vspace{-0.5cm}

\begin{figure}[t!]
\centering
\includegraphics[scale=0.39]{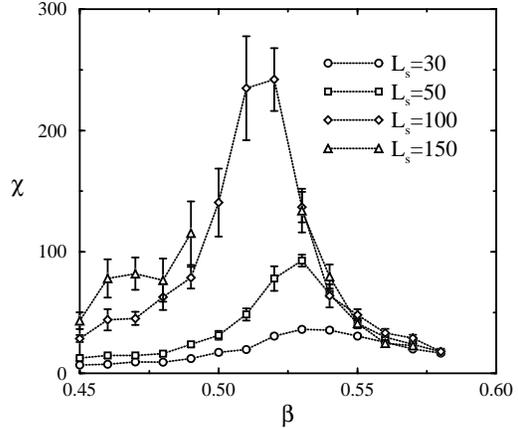}
\vspace{-1.3cm}
\caption{\small Susceptibility $\chi$ vs. $\beta$.}
\label{fig:suscept}
\end{figure}
%\vspace{-0.5cm}

In order to study the behaviour of the chiral symmetry
at $T\!>\!0$ in the absence of a fermion bare mass in the Lagrangian
the best thing to measure is
an effective ``order parameter'' $|\Phi| \equiv \sqrt{\Sigma^2 + \Pi^2}$, 
which is a projection onto
the direction of $\Phi^{\alpha}  \equiv (\Sigma, \Pi)$ separately for 
each configuration. 
%\begin{figure}[htb]
%\centerline{ \epsfysize=2.4in
%\setlength\epsfxsize{200pt}
%         \epsfbox{fig2.eps}}
%\caption{Order parameter $\vert\Phi\vert$ vs. $\beta$ for $L_t=4$
%for both $U(1)$ and $Z_2$ symmetric models.}
%\label{fig:order_para}
%\end{figure}
%\begin{figure}[tbh]
%                \centerline{ \epsfysize=2.4in
%                             \epsfbox{fig3.eps}}
%
%\caption{Susceptibility $\chi$ vs. $\beta$.}
%\label{fig:suscept}
%\end{figure}

In Fig.\ref{fig:order_para} we plot $|\Phi|$ versus $\beta \! \equiv \! 1/g^2$
for different lattice sizes together with the results from simulations
of the model with a $Z_2$ chiral symmetry. 
At the bulk critical coupling
$\beta_c^{bulk}\! \equiv 1/g^2 \! \approx \! 0.86$ \cite{strouthos01}
the lattice spacing becomes zero and $T \! \rightarrow \! \infty$.
It is clear that
the order parameter of the $Z_2$ model is independent of the lattice size until just before the
transition at $\beta \!=\! \beta_c^{Z_2} \!= \! 0.565(3)$,
whereas in the $U(1)$ model $|\Phi|$ has a strong size dependence for a large range of values
of $\beta$, i.e. it
decreases rapidly as the spatial volume increases in accordance with the expectation that chiral
symmetry should be restored for $T\!>\!0$.
The finite spatial extent $L_s$ provides a cut-off for the divergent correlation length and according to
the BKT scenario
the slow decay of the correlation function $\langle e^{i\theta(x)} e^{-i\theta(0)} \rangle$
with exponent $\eta(T)\!<\!0.25$ for $T\!<\!T_{BKT}$ ensures a non-zero magnetization 
even in a system with very large size.
%\begin{figure}[p]
%
%                \centerline{ \epsfysize=2.3in
%                             \epsfbox{fig3.eps}}
%\caption[]{Susceptibility $\chi$ vs. $\beta$.}
%\label{fig:suscept}
%\end{figure}
In Fig.\ref{fig:suscept} we plot the susceptibility of the order parameter
$\chi = V (\langle |\Phi|^2 \rangle - \langle |\Phi| \rangle^2)$, measured on 
lattices with different spatial size. We observe from this figure that:
(a) there is a phase transition since the peak of $\chi$ diverges
as we increase $L_s$; (b) the transition occurs at a much smaller temperature
than the critical temperature of the $Z_2$-symmetric model, because 
the IR fluctuations are stronger in the continuous symmetry case and (c)
in the low $T$ phase the susceptibility has a a stronger size dependence 
than in the high $T$ phase and the errors at low $T$ are much larger than 
at high $T$. This is strong evidence that the system in critical 
in the low $T$ phase in accordance with the BKT scenario.
%\begin{figure}[htb]
%                \centerline{ \epsfysize=2.4in
%                             \epsfbox{fig4.eps}}
%\caption{Specific heat $C_v$ vs. $\beta$.}
%\label{fig:spec_heat}
%\end{figure}
We also measured the specific heat $C_V$, 
which we calculated from the
fluctuations of the bosonic action
$S_b = \frac{1}{2} \sum_x [\sigma^2(x) + \pi^2(x)]$.
$C_v$ is given by
%\begin{equation}
$C_v = \frac{\beta^2}{V} (\langle S_b^2 \rangle - \langle S_b \rangle^2)$.
%\end{equation}
%and is plotted versus coupling for $L_s=50,100,150$ in
%Fig.\ref{fig:spec_heat}.
We observed that $C_v$ has a broad peak at $\beta \simeq 0.50$ and 
it does not show any divergent behaviour or significant finite size effects.
This is also consistent with the BKT scenario, according to which $\alpha\!=\!-2$.
%\begin{figure}[p]
%                \centerline{ \epsfysize=2.3in
%                             \epsfbox{fig5.eps}}
%\caption[]{Fermion screening mass $M_s$ vs. $\beta$. The horizontal line
%shows the lowest Matsubara mode $M_0^{lat}$.}
%\label{fig:mass}
%\end{figure}

\begin{figure}[t!]
\centering
\includegraphics[scale=0.39]{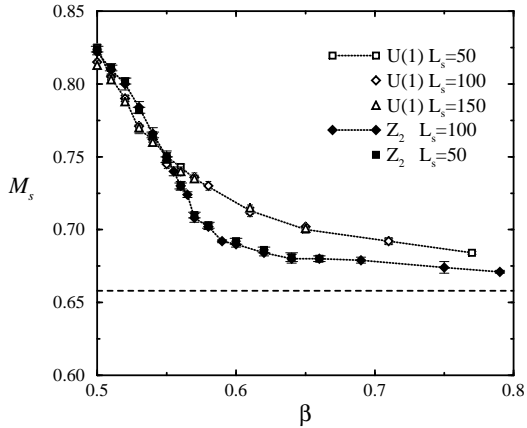}
\vspace{-1.3cm}
\caption{\small Fermion screening mass $M_s$ vs. $\beta$. The horizontal line
shows the lattice lowest Matsubara mode $\omega_0$.}
\label{fig:mass}
\end{figure}

Finally, we measured the fermion screening mass $M_s(T)$ from the exponential decay 
of the spatial correlator along one of the spatial lattice axes.
The results extracted from the simulations
of the $U(1)$ and $Z_2$ models on lattices with different $L_s$,
are shown in Fig.\ref{fig:mass}. 
If we assume that the temperature effects are absorbed into a temperature dependent
pole mass $M(T)$ and a ``renormalization'' of the speed of light coefficient
$A(T)$ then we have $M_s^2(T)\!=\!A(T)[M^2(T)+ \omega_0^2]$.
We can infer from Fig.\ref{fig:mass} that the screening mass is independent of $L_s$
for almost all values of $\beta$. 
In the $Z_2$ case we ascribe mass generation to orthodox chiral symmetry
breaking.
The similarity of $M_s^{U(1)}$ and $M_s^{Z_2}$ at low $T$ where $A \approx 1$
suggests that the pole mass $M(T)$ is nonzero.
By comparing Fig.\ref{fig:suscept} with Fig.\ref{fig:mass} we infer
that the fermions are massive
in the phase where both chiral symmetry is restored and
the system is not critical.
This implies the existence of a pseudogap phase for
$\beta_{BKT} \leq \beta \leq \beta_c^{Z_2}$.
%\vspace{-0.5cm}
The deviation of $M_s(T)$ from $\omega_0$ for $\beta\! >\! 0.56$
could be attributted to either $A(T)\!>\!1$, $M(T)\!>\!0$ or
discretization 
effects. To understand the physics in this regime
these effects need to be disentangled via a detailed study of the 
fermion dispersion relation closer to the continuum limit on lattices with 
$L_t\! \geq \!8$. For this reason we cannot at this stage identify a second
phase transition where mass generation switches off.

%\vspace{-0.5cm}

\section{Summary and Outlook}
The numerical results presented in this report provide strong 
evidence that the nonzero temperature phase structure
of the $U(1)$-symmetric GNM$_3$ is consistent with the BKT scenario. 
We have also shown that there is a pseudogap phase i.e., the fermion 
mass remains nonzero above $T_{BKT}$.
We are currently extending our work in various directions. We wish 
(i) to increase our statistics in order to extract the exponents 
$\eta(T)$; (ii) to study the fermion dispersion 
relation and (iii) to understand the $N_f$ dependence of our various results.  

\section*{Acknowledgements}
This project is being done in collaboration with Simon Hands
and John Kogut. Costas Strouthos is supported by a Leverhulme Trust grant.

\end{document}